\documentclass[aps,prd,preprintnumbers,amsmath,amssymb,nofootinbib,11pt]{revtex4}
\usepackage{eepic}
\usepackage{indentfirst}
\usepackage{mathrsfs}
\usepackage{fancyhdr}
\usepackage{ulem}
\usepackage{float}
\usepackage{graphicx}
\usepackage[
colorlinks=true, linkcolor=black, breaklinks=true, urlcolor=blue,
citecolor=green]{hyperref}
\usepackage{epstopdf}
\usepackage{bm,bbm}

\usepackage{xcolor}

\usepackage{tabularx}
\usepackage{subfigure}
\usepackage{epsfig}
\usepackage{color}
\usepackage{slashed}
\usepackage{hyperref}

\newcommand{\be}{\begin{equation}}
\newcommand{\ee}{\end{equation}}
\renewcommand{\L}{\mathscr{L}}

\newcommand{\bra}{\langle}
\newcommand{\ket}{\rangle}
\newcommand{\nn}{\nonumber}
\newcommand{\MeV}{\,\text{MeV}}
\newcommand{\GeV}{\,\text{GeV}}
\renewcommand{\vec}[1]{\mathbf{#1}}
\newcommand{\diff }{{\text{d}}}

\begin{document}
\pagestyle{plain}

\title {\boldmath $\phi(2170)$ decaying to $\phi \pi\pi$ and $\phi K\bar{K}$}
\author{ Yun-Hua~Chen}\email{yhchen@ustb.edu.cn}
\affiliation{School of Mathematics and Physics, University
of Science and Technology Beijing, Beijing 100083, China}

\begin{abstract}

Within the framework of dispersion theory, we study the the processes
$e^+e^-\to \phi(2170) \to \phi \pi\pi(K\bar{K})$.
The strong pion--pion final-state interactions, especially
the $K\bar{K}$ coupled channel in the $S$-wave,
are taken into account in a model-independent way using the Omn\`es function solution.
Through fitting the experimental data of the $\pi\pi$ and $\phi\pi$ invariant mass distributions of the $e^+e^- \to \phi(2170) \to \phi \pi^+\pi^-$ process,
the low-energy constants in the chiral Lagrangian are determined.
The theoretical prediction for the cross sections ratio ${\sigma(e^+e^- \to \phi(2170)\to \phi K^+ K^-)}/{\sigma(e^+e^- \to \phi(2170)\to \phi\pi^+\pi^-)}$ is given, which could be useful for
selecting the physical solution when the fit to the $e^+e^- \to \phi K^+ K^-$ cross section distribution is available in the future.
Our results suggest that above the kinematical threshold of $\phi K\bar{K}$, the mechanism $e^+e^- \to \phi K^+ K^-$ with the kaons rescattering to a pion pair plays an important role in the $e^+e^- \to \phi\pi^+\pi^-$ transition.

\end{abstract}

\maketitle

\newpage
\section{Introduction}

The vector strangeoniumlike state $\phi(2170)$ was first discovered in 2006 by the BaBar Collaboration in the initial-state radiation process
$e^+e^-\to \gamma_{ISR} \phi f_0(980)$~\cite{Aubert:2006bu,Aubert:2007ur,Aubert:2007ym,Lees:2011zi}, and later confirmed by BESII~\cite{Ablikim:2007ab}, Belle~\cite{Shen:2009zze}, and BESIII~\cite{Ablikim:2014pfc,Ablikim:2017auj} collaborations. Its mass and width were measured to be $M = 2188 \pm 10$ MeV and $\Gamma = 83 \pm 12$ MeV respectively, and its spin-parity quantum number is $J^{PC} = 1^{--}$~\cite{ParticleDataGroup:2022pth}. The nature of $\phi(2170)$ has remained controversial, and models have been proposed to interpret the
$\phi(2170)$ as a hybrid
state~\cite{Ding:2006ya}, an excited
strangeonium~\cite{Ding:2007pc}, a hidden-strangeness baryon-antibaryon state ($q q s \bar q \bar q \bar s$)~\cite{Abud:2009rk}, a bound states of $\Lambda \bar \Lambda(^3S_1)$~\cite{Zhao:2013ffn}, a tetraquark
state~\cite{Wang:2006ri,Chen:2008ej,Ali:2011qi,Chen:2018kuu,Ke:2018evd,Liu:2020lpw}, and a dynamically generated state in the $\phi f_0(980)$ system~\cite{AlvarezRuso:2009xn,Coito:2009na} or the $\phi K \bar K$ system~\cite{Napsuciale:2007wp,GomezAvila:2007ru,MartinezTorres:2008gy,Malabarba:2023zez}.

\begin{figure}
\centering
\includegraphics[height=6cm,width=10cm]{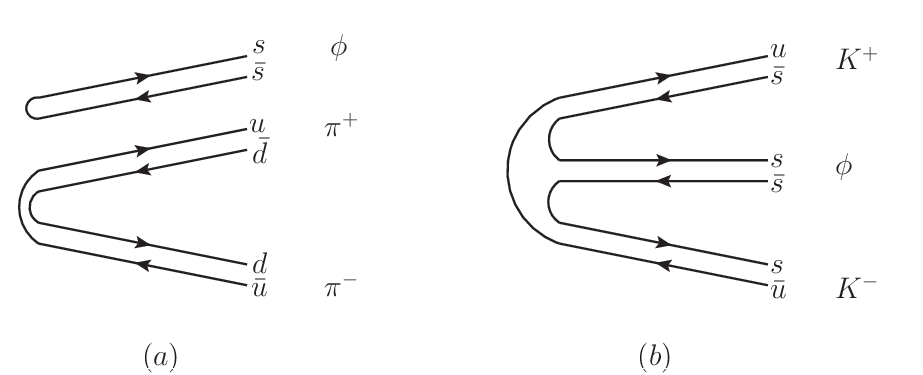}
\caption{Quark diagrams of the final states of $ \phi \pi \pi$ and $\phi K \bar{K}$.
}\label{fig.QuarkDiagramofFinalStates}
\end{figure}

\begin{figure}
\centering
\includegraphics[height=6cm,width=8cm]{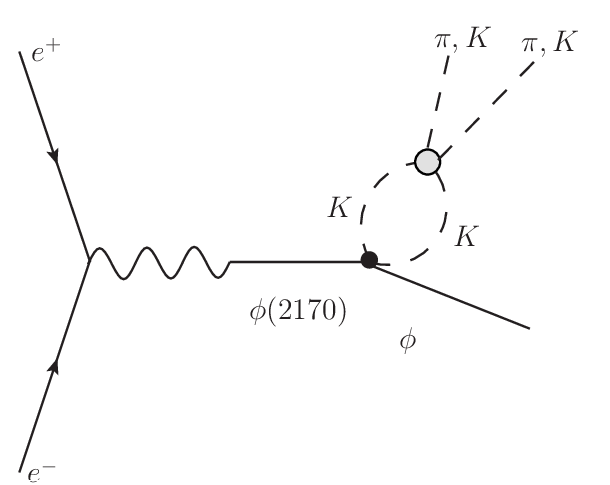}
\caption{Feynman diagrams considered for $e^+ e^- \to
\phi(2170) \to \phi \pi \pi (\phi K \bar{K})$. The gray blob denotes the effects of FSI.
}\label{fig.FeynmanDiagram}
\end{figure}

Since both $\phi(2170)$ and $Y(4230)$ are observed in $e^+ e^-$ annihilation through initial state radiation, $\phi(2170)$ is often taken as the strange analogue of $Y(4230)$. Similar to the observation of $Z_c(3900)$ in the $ J/\psi \pi$ invariant mass spectrum in $ Y(4230) \to J/\psi \pi\pi$ process, recently the BESIII Collaboration searched for a strangeoniumlike structure $Z_s$ decaying into $\phi\pi$ in the
$\phi(2170) \to \phi \pi\pi$ process~\cite{BESIII:2018rdg}. No $Z_s$ signal was observed in the $\phi\pi$ invariant mass spectrum. On the other hand, the Born cross sections for the channel $e^+e^- \to \phi K^+K^-$ have been measured for the first time at center-of-mass energies between 2.100 and 3.080 GeV~\cite{BESIII:2019ebn}. In this work, we will study the $\pi\pi$ and $\phi\pi$ invariant mass spectrum of
$e^+e^- \to \phi(2170) \to \phi \pi^+\pi^-$ process, and the ratio of cross sections ${\sigma(e^+e^- \to \phi(2170)\to \phi K^+ K^-)}/{\sigma(e^+e^- \to \phi(2170)\to \phi\pi^+\pi^-)}$.
As shown in Fig.~\ref{fig.QuarkDiagramofFinalStates}, the quark lines of the $\phi$ and $\pi\pi$ final states are disconnected and thus at tree level the leading electromagnetic contributions to the $e^+e^- \to \phi \pi^+\pi^-$ process from the exchange of a virtual photon is suppressed~\cite{Napsuciale:2007wp}. The $\pi\pi$ invariant mass goes up to more than $1.1\GeV$, and in this energy region there are strong coupled-channel final-state interactions (FSIs) especially in the $S$-wave.
In this work, we will take into account the strong FSIs model-independently using dispersion theory, and study the contribution of the mechanism $e^+e^- \to \phi K^+ K^-$ with the kaons rescattering to a pion pair to $e^+e^- \to \phi\pi^+\pi^-$ transition.
At low energies, the amplitude should agree with the leading chiral results,
therefore the
subtraction terms in the dispersion relations can be determined by matching to the chiral contact terms.
For the leading contact couplings for $\phi(2170)\phi PP$, here $P$ denotes the pseudoscalar meson $\pi$ or $K$, we construct the chiral Lagrangians in the spirit of
the chiral effective field theory ($\chi$EFT)~\cite{Mannel}. The parameters are then fixed by fitting to the BESIII data.
The relevant Feynman diagrams considered is given in Fig.~\ref{fig.FeynmanDiagram}.

This paper is organized as follows. In Sec.~\ref{theor}, we present
the theoretical framework and the calculation of the
amplitudes as well as the dispersive treatment of the FSI. In Sec.~\ref{pheno},
we fit the experimental data of the $\pi\pi$ and $\phi\pi$
invariant mass distribution to determine the coupling
constants, and discuss the contribution of $e^+e^- \to \phi K^+ K^-$ with the kaons rescattering to a pion pair to the $e^+e^- \to \phi\pi^+\pi^-$ transition.
A summary is given in Sec.~\ref{conclu}.

\section{Theoretical framework}\label{theor}
\subsection{Lagrangians}

The $\phi$ meson can be decomposed into
SU(3) singlet and octet components of light quarks,
\be
\label{eq.phiComponents} |\phi\rangle=s\bar{s}=\frac{\sqrt{3}}{3}|V_1\rangle-\frac{\sqrt{6}}{3}|V_8\rangle\,,
\ee
where $|V_1\rangle \equiv
\frac{1}{\sqrt{3}}(u\bar{u}+d\bar{d}+s\bar{s})$ and $|V_8\rangle \equiv
\frac{1}{\sqrt{6}}(u\bar{u}+d\bar{d}-2
s\bar{s})$.
In the $\phi(2170) \to \phi PP$ transition, the two pseudoscalars in the final state must come from light-flavor sources.
There are two types of sources that the two pseudoscalars may come from, one is the possible light-quark components contained in the $\phi(2170)$ (e.g. in the $\phi f_0(980)$ molecule or the tetraquark scenarios), and the other possibility is that the two pseudoscalars are excited by the $\phi(2170)$ from vacuum (e.g. in the pure $s \bar{s}$ or the hybrid state scenarios).
In our study we do not distinguish these two types of pseudoscalars sources but take into account them in an unified scheme, and call them both are ``provided'' by the $\phi(2170)$.
If the $\phi(2170)$
contains no $u, d$ quarks (as in the pure $s \bar{s}$ or the hybrid state scenarios), the light-flavor sources excited by the $\phi(2170)$ from vacuum has
to be in the form of an SU(3) singlet state. While since the structure of the $\phi(2170)$ has remained controversial, the relative strengths between the light-flavor SU(3) singlet part and SU(3) octet part acting in this transition is undetermined.
Therefore, considering the two pseudoscalars sources provided by the $\phi(2170)$ in the $\phi(2170) \to \phi PP$ transition, the $\phi(2170)$ can be decomposed into SU(3) singlet and octet components of light quarks,
\be
\label{eq.YComponents} |\phi(2170)\rangle=a|Y_1\rangle+b|Y_8\rangle\,.
\ee
The values of the component strengths $a$ and $b$ can not be determined in this study, since they always appear in the combination of $(a-\sqrt{2}b)$ in the chiral contact amplitude for $\phi(2170) \to \phi PP$ transition that will be given in subsection~\ref{subsectionB}.
Expressed in terms of a $3\times3$ matrix in the SU(3) flavor space, it is written as
\begin{equation}
 \frac{a}{\sqrt{3}} Y_1 \cdot \mathbbm{1}+\frac{b}{\sqrt{6}} Y_8\cdot \text{diag} \left( 1,  1, - 2\right)    .
\end{equation}

The effective Lagrangian for the $\phi(2170)\phi\pi\pi$ and
$\phi(2170)\phi K\bar{K}$ contact couplings, at leading order in the chiral expansion,
reads~\cite{Mannel,Chen:2019mgp}
\begin{align}\label{LagrangianYphipipi}
\L_{\phi(2170)\phi PP} =& g_1\bra Y_{1}^\mu V_{1\mu} \ket \bra u_\nu
u^\nu\ket -\sqrt{2}g_1\bra  Y_1^\mu \ket \bra V_{8\mu} u_\nu u^\nu\ket +g_8\bra V_{1\mu}  \ket \bra Y_{8}^\mu u_\nu
u^\nu\ket -\sqrt{2}g_8 \bra Y_{8}^\mu V_{8\mu} u_\nu u^\nu\ket\nonumber\\&
+h_1\bra Y_{1}^\mu V_{1\nu} \ket \bra u_\mu
u^\nu\ket -\sqrt{2}h_1\bra  Y_1^\mu \ket \bra V_{8\nu} u_\mu u^\nu\ket +h_8\bra V_{1\nu}  \ket \bra Y_{8}^\mu u_\mu
u^\nu\ket -\sqrt{2}h_8 \bra Y_{8}^\mu V_{8\nu} u_\mu u^\nu\ket
+\mathrm{H.c.}\,,
\end{align}
where $\langle\ldots\rangle$ denotes the trace in the SU(3) flavor space. In Eq.~\eqref{LagrangianYphipipi}, the Lagrangian are constructed by placing the SU(3) singlet parts and the SU(3) octet parts into different SU(3) flavor traces.  The SU(3) octet of the pseudo-Goldstone bosons from the spontaneous breaking of chiral symmetry, can
be filled nonlinearly into
\begin{equation}
u_\mu = i \left( u^\dagger\partial_\mu u\, -\, u \partial_\mu
u^\dagger\right) \,, \qquad
u = \exp \Big( \frac{i\Phi}{\sqrt{2}F} \Big)\,,
\end{equation}
with the Goldstone fields
\begin{align}
\Phi &=
 \begin{pmatrix}
   {\frac{1}{\sqrt{2}}\pi ^0 +\frac{1}{\sqrt{6}}\eta _8 } & {\pi^+ } & {K^+ }  \\
   {\pi^- } & {-\frac{1}{\sqrt{2}}\pi ^0 +\frac{1}{\sqrt{6}}\eta _8} & {K^0 }  \\
   { K^-} & {\bar{K}^0 } & {-\frac{2}{\sqrt{6}}\eta_8 }  \\
 \end{pmatrix} . \label{eq:u-phi-def}
\end{align}
Here $F$ is the pion decay constant in the chiral limit, and we take the physical value $92.1\MeV$ for it.

The gauge-invariant $\gamma^\ast(\mu)$ and $\phi(2170) (\nu)$ coupling is given by
\be
iV_{\gamma^{\ast\mu}Y^\nu}=2i(g^{\mu\nu}p^2-p^\mu
p^\nu)c_\gamma \,,
\ee
where $p$ is the momentum of the virtual photon $\gamma^\ast.$

\subsection{Amplitudes of \boldmath{$ \phi(2170) \to \phi PP $} processes} \label{subsectionB}

The decay amplitude of $ \phi(2170)(p_a) \to
\phi(p_b) P(p_c)P(p_d) $ can described in terms of the
Mandelstam variables
\begin{align}
s &= (p_c+p_d)^2 , \qquad
t_P=(p_a-p_c)^2\,, \qquad u_P=(p_a-p_d)^2\,,\nn\\
3s_{0P}&\equiv s+t_P+u_P=
 M_{\phi(2170)}^2+M_{\phi}^2+2m_P^2  \,.
\end{align}
The variables $t_P$ and $u_P$ can
be expressed in terms of $s$ and the scattering angle $\theta$
according to
\begin{align}
t_P &= \frac{1}{2} \left[3s_{0P}-s+\kappa_P(s)\cos\theta \right]\,,&
u_P &= \frac{1}{2} \left[3s_{0P}-s-\kappa_P(s)\cos\theta \right]\,, \nn\\
\kappa_P(s) &\equiv \sigma_P
\lambda^{1/2}\big(M_{\phi(2170)}^2,M_{\phi}^2,s\big) \,, & \sigma_P &\equiv
\sqrt{1-\frac{4m_P^2}{s}} \,, \label{eq:tu}
\end{align}
where $\theta$ is defined as the angle between the positive
pseudoscalar meson and the $\phi(2170)$ in the rest frame of the $PP$ system, and
$\lambda(a,b,c)=a^2+b^2+c^2-2(ab+ac+bc)$ is the K\"all\'en triangle function. We define $\vec{q}$ as the
3-momentum of final $\phi$ in the rest frame of the $\phi(2170)$ with
\be \label{eq:q} |\vec{q}|=\frac{1}{2M_{\phi(2170)}}
\lambda^{1/2}\big(M_{\phi(2170)}^2,M_{\phi}^2,s\big) \,. \ee

Using the Lagrangians in
Eq.~\eqref{LagrangianYphipipi}, we can calculate the chiral contact terms for $\phi(2170) \to \phi \pi^+\pi^- $ and $\phi(2170)\to \phi K^+ K^- $ processes
\begin{align}
M^{\chi,\pi}(s,\cos\theta)&=0\,, \notag\\
\label{eq.ContactPi+KRaw}
M^{\chi,K}(s,\cos\theta)&=-\frac{3}{F^2}\bigg[2\Big(g_1-\sqrt{2}g_8\Big) p_c\cdot
p_d \epsilon_Y\cdot\epsilon_\phi +\Big(h_1-\sqrt{2}h_8\Big)\Big(p_c\cdot\epsilon_Y p_d\cdot\epsilon_\phi+p_c\cdot\epsilon_\phi p_d\cdot\epsilon_Y\Big) \bigg]\,.
\end{align}
Notice that the quark lines of the $\phi$ and $\pi\pi$ final states are disconnected and therefore at tree level the leading electromagnetic contributions to the $e^+e^- \to \phi \pi^+\pi^-$ process from the exchange of a virtual photon is suppressed~\cite{Napsuciale:2007wp}. The amplitude $M^{\chi,\pi}(s,\cos\theta)=0$ in Eq.~\eqref{eq.ContactPi+KRaw} agrees with this observation. Thus the mechanism $e^+e^- \to \phi K\bar{K}$ with the kaons rescattering to a pion pair may be an important contribution to $e^+e^- \to \phi \pi^+\pi^-$.

The appropriate helicity amplitudes $M_{\lambda_1\lambda_2}^{\chi,\pi(K)}(s,\cos\theta)$, with $\lambda_1$($\lambda_2$) denoting the $\phi(2170)$($\phi$) helicities respectively, are obtained by inserting explicit expressions for
the polarization vectors $\epsilon^\mu(p_i,\lambda)$ occurring in the amplitudes Eq.~\eqref{eq.ContactPi+KRaw}, taken from Ref.~\cite{Lutz:2011xc},
\begin{eqnarray}\label{eq.polarizationvectors}
&& \epsilon^\mu(p_{\phi(2170)}, \pm 1)=\left(
                           \begin{array}{c}
                             0 \\
                              \frac{\mp\cos \theta }{\sqrt{2}} \\
                            \frac{- i}{\sqrt{2}} \\
                              \frac{ \pm \sin \theta }{\sqrt{2}} \\
                           \end{array}
                         \right)\,,\;
\epsilon^\mu(p_{\phi(2170)},  0)=      \left(
                           \begin{array}{c}
                             \frac{\vec{q}}{M_{\phi(2170)}} \\
                             \frac{\sqrt{M_{\phi(2170)}^2+\vec{q}^2}}{M_{\phi(2170)}} \sin \theta \\
                             0 \\
                             \frac{\sqrt{M_{\phi(2170)}^2+\vec{q}^2}}{M_{\phi(2170)}} \cos \theta \\
                           \end{array}
                         \right)\,,       \nonumber
\nonumber\\
&& \epsilon^\mu(p_\phi, \pm 1) = \left(
                           \begin{array}{c}
                             0 \\
                            \frac{\pm \cos \theta }{\sqrt{2}} \\
                            \frac{-i}{\sqrt{2}} \\
                            \frac{\mp \sin \theta }{\sqrt{2}} \\
                           \end{array}
                         \right) \,, \;
\epsilon^\mu(p_\phi, 0) = \left(
                           \begin{array}{c}
                             \frac{\vec{q}}{M_\phi} \\
                             -\frac{\sqrt{M_\phi^2+\vec{q}^2}}{M_\phi} \sin \theta \\
                             0 \\
                             -\frac{\sqrt{M_\phi^2+\vec{q}^2}}{M_\phi} \cos \theta \\
                           \end{array}
                         \right)\,.
\end{eqnarray}
Note in this study we need to perform the partial-wave projections of the $PP$ system to take into account the final-state interactions. We can analytically continue the $\phi(2170) \to \phi PP$ decay amplitude
to the $\phi(2170)\phi\to PP$ scattering amplitude, since the partial-wave decomposition to the latter is easier. Therefore, the expressions for the polarization vectors given in Eq.~\eqref{eq.polarizationvectors} are defined in
the $PP$ rest frame. 

The partial-wave projection of the $\phi(2170) \to \phi PP$ helicity amplitudes is given
\begin{equation}\label{eq.partialwaveamp}
M_{\lambda_{1} \lambda_{2}}^{\chi,\pi(K),l}(s)=\frac{2 l+1}{2} \int \mathrm{d} \cos \theta d_{\lambda_1-\lambda_2, 0}^{l}(\theta) M_{\lambda_{1} \lambda_{2}}^{\chi,\pi(K)}(s,\cos\theta)\,,
\end{equation}
where  $d_{\lambda_1-\lambda_2,0}^l(\theta)$ are the small Wigner-d functions.

\subsection{Final-state interactions with a dispersive approach, Omn\`es solution }

The strong FSIs between two pseudoscalar mesons can be taken account of
model-independently using dispersion theory. Since the invariant
mass of the pion pair reaches above the $K\bar{K}$ threshold, we
will take into account the coupled-channel ($\pi\pi$ and $K\bar
K$) FSI for the dominant $S$-wave component, while for the $D$-wave
only the single-channel FSI will be considered.
Similar methods to consider the FSI have been applied previously e.g.\ in
Refs.~\cite{Moussallam-gamma,KubisPlenter,ZHGuo,Kang,Dai:2014lza,Dai:2014zta,Dai:2016ytz,Chen2016,Chen:2016mjn,Chen:2019gty,Chen:2019mgp,Chen:2021aud}.

For $\phi(2170) \to \phi PP$, the
partial-wave decomposition of the helicity amplitude including FSIs reads
\be
M^{P,\text{decay}}(s,\cos\theta)  = \sum_{\lambda_1 \lambda_2}\sum_{l=0}^{\infty} M_{\lambda_1\lambda_2}^{P,l}(s) d_{\lambda_1-\lambda_2,0}^l(\theta)\,.
\label{eq.PartialWaveFullAmplitude}\ee

For the $S$-wave, we will take into account the two-channel
rescattering effects. The two-channel unitarity condition reads
\begin{equation}\label{eq.unitarity2channel}
\textrm{disc}\, \vec{M}_{\lambda_1 \lambda_2}^0(s)=2i T_0^{0\ast}(s)\Sigma(s)
\vec{M}_{\lambda_1 \lambda_2}^0(s),
\end{equation}
where the two-dimensional vectors $\vec{M}_{\lambda_1 \lambda_2}^0(s)$ contains both the $\pi\pi$ and the $K\bar{K}$ final states,
 \begin{equation}
\vec{M}_{\lambda_1 \lambda_2}^0(s)=\left( {\begin{array}{*{2}c}
   {M_{\lambda_1 \lambda_2}^{\pi,0}(s)} \\
   {\frac{2}{\sqrt{3}}M_{\lambda_1 \lambda_2}^{K,0}(s)}  \\\end{array}} \right).
 \end{equation}
The two-dimensional matrices $T_0^0(s)$ and $\Sigma(s)$ are represented as
\begin{equation}\label{eq.T00}
T_0^0(s)=
 \left( {\begin{array}{*{2}c}
   \frac{\eta_0^0(s)e^{2i\delta_0^0(s)}-1}{2i\sigma_\pi(s)} & |g_0^0(s)|e^{i\psi_0^0(s)}   \\
  |g_0^0(s)|e^{i\psi_0^0(s)} & \frac{\eta_0^0(s)e^{2i\left(\psi_0^0(s)-\delta_0^0(s)\right)}-1}{2i\sigma_K(s)} \\
\end{array}} \right),
\end{equation}
and $\Sigma(s)\equiv \text{diag}
\big(\sigma_\pi(s)\theta(s-4m_\pi^2),\sigma_K(s)\theta(s-4m_K^2)\big)$.
There are three input functions in the $T_0^0(s)$ matrix: the
$\pi\pi$ $S$-wave isoscalar phase shift $\delta_0^0(s)$, and the modulus and phase of the $\pi\pi \to
K\bar{K}$ $S$-wave amplitude $g_0^0(s)=|g_0^0(s)|e^{i\psi_0^0(s)}$. We will use the parametrization of the $T_0^0(s)$ matrices given in Refs.~\cite{Leutwyler2012,Moussallam2004}.
Note that the relation between the  inelasticity parameter $\eta_0^0(s)$ in Eq.~\eqref{eq.T00} and the modulus $|g_0^0(s)|$
\begin{equation}
\eta_0^0(s)=\sqrt{1-4\sigma_\pi(s)\sigma_K(s)|g_0^0(s)|^2\theta(s-4m_K^2)}\,.
\end{equation}
These inputs are used up to $\sqrt{s_0}=1.3\GeV$, and above $s_0$ the $f_0(1370)$ and $f_0(1500)$ resonances coupling strongly to
$4\pi$ will contribute further inelasticities~\cite{Tanabashi:2018oca,Ropertz:2018stk}. Above $s_0$, we guide smoothly the phases
$\delta_0^0(s)$ and $\psi_0^0$ to 2$\pi$ by
means of~\cite{Moussallam2000}
\begin{equation}
\delta(s)=2\pi+(\delta(s_0)-2\pi)\frac{2}{1+({s}/{s_0})^{3/2}}\,.
\end{equation}

\noindent The solution of the coupled-channel unitarity condition in
Eq.~\eqref{eq.unitarity2channel} is given by
\begin{equation}\label{OmnesSolution2channel}
\vec{M}_{\lambda_1 \lambda_2}^0(s)=\Omega(s)\vec{P}^{n-1}(s)
\,,
\end{equation}
where $\Omega(s)$ satisfies the homogeneous coupled-channel
unitarity relation
\begin{equation}\label{eq.unitarity2channelhomo}
\textrm{Im}\, \Omega(s)=T_0^{0\ast}(s)\Sigma(s) \Omega(s),
\hspace{1cm}  \Omega(0)=\mathbbm{1} \,,
\end{equation}
and its numerical results have been computed, e.g., in
Refs.~\cite{Leutwyler90,Moussallam2000,Hoferichter:2012wf,Daub}.

For the $D$-wave, the single-channel FSI will
be considered. In the elastic $PP$ rescattering region,
the partial-wave unitarity condition is
\begin{equation}\label{eq.unitarity1channel}
\textrm{Im}\, M_{\lambda_1 \lambda_2}^{P,2}(s)= M_{\lambda_1 \lambda_2}^{P,2}(s)
\sin\delta_2^0(s) e^{-i\delta_2^0(s)}\,,
\end{equation}
where the phase of the $D$-wave isoscalar amplitude
$\delta_2^0$ coincides with
the $PP$ elastic phase shift, as required by
Watson's theorem~\cite{Watson1,Watson2}.
The Omn\`es solution of Eq.~\eqref{eq.unitarity1channel}
reads
\be\label{OmnesSolution1channel}
M_{\lambda_1 \lambda_2}^{P,2}(s)=\Omega_2^0(s)P_2^{n-1}(s) \,, \ee
where the polynomial $P_2^{n-1}(s)$ is a subtraction function, and the
Omn\`es function is defined as~\cite{Omnes}
\begin{equation}\label{Omnesrepresentation}
\Omega_2^0(s)=\exp
\bigg\{\frac{s}{\pi}\int^\infty_{4m_\pi^2}\frac{\diff x}{x}
\frac{\delta_2^0(x)}{x-s}\bigg\}\,.
\end{equation}
We will use the result of $\delta_2^0(s)$ given in Ref.~\cite{Pelaez}, which is smoothly continued to $\pi$ for $s\to\infty$.

On the other hand, at
low energies the partial-wave amplitudes $\vec{M}_{\lambda_1 \lambda_2}^0(s)$ and $M_{\lambda_1 \lambda_2}^2(s)$ should match to those from $\chi$EFT. Namely, if one switches off the FSI with $s=0$, $\Omega(0)=\mathbbm{1}$ and $\Omega_2^0(0)=1$,
the subtraction functions should agree well with the low-energy chiral amplitudes
given in Eq.~\eqref{eq.partialwaveamp}.
Thus, for the
$S$-wave, the integral equation takes the form
\begin{equation}\label{M02channel}
\vec{M}_{\lambda_1 \lambda_2}^0(s)=\Omega(s)\vec{M}_{\lambda_1 \lambda_2}^{\chi,0}(s)
\,,
\end{equation}
where $ \vec{M}_{\lambda_1 \lambda_2}^{\chi,0}(s)=\big(
   M_{\lambda_1 \lambda_2}^{\chi,\pi,0}(s),
   2/\sqrt{3}\,M_{\lambda_1 \lambda_2}^{\chi,K,0}(s)
   \big)^{T}$, while
for the $D$-wave, it reads
\be\label{M21channel}
M_{\lambda_1 \lambda_2}^{P,2}(s)=\Omega_2^0(s)M_{\lambda_1 \lambda_2}^{\chi,P,2}(s) \,.
\ee
Note that as given in Eq.~\eqref{eq.ContactPi+KRaw}, the chiral contact amplitudes for $\phi(2170) \to \phi \pi^+\pi^- $ process $M_{\lambda_1 \lambda_2}^{\chi,\pi,0(2)}(s)$
equal 0,
and only the non-zero amplitudes $M_{\lambda_1 \lambda_2}^{\chi,K,0(2)}(s)$ affect the numerical calculation. 

The polarization-averaged modulus-square of the $e^+e^- \to \phi(2170)
\to \phi \pi^+\pi^-$ amplitude can be written as
\begin{equation}
|\bar{M}(E^2,s,\cos\theta)|^2 = \frac{4\pi\alpha
c_\gamma^2|M^{\pi,\text{decay}}(s,\cos\theta)|^2}{3|E^2-M_{\phi(2170)}^2+iM_{\phi(2170)}\Gamma_{\phi(2170)}|^2
M_\phi^2}\left[  8 M_\phi^2 E^2+(s-E^2-M_\phi^2)^2
\right],\label{eq.eetoJpsipipiAmplitudeSquar}
\end{equation}
where $E$ is the center-of-mass energy of the
$e^+e^-$ system, and we set the $\gamma^\ast \phi(2170)$ coupling
constant $c_\gamma$ to 1 since it can be absorbed into the overall
normalization in the fitting of the event distributions. Here we use the energy-independent width for the
$\phi(2170)$, and the values of the $\phi(2170)$ mass is taken as $2125\MeV$, which is the center-of-mass energy measured by BESIII detector in Ref.~\cite{BESIII:2018rdg}.
The width of $\phi(2170)$ is taken as $100\MeV$ from PDG~\cite{ParticleDataGroup:2022pth}.

The $\pi\pi$ invariant mass distribution of $e^+e^- \to \phi \pi^+\pi^-$ reads
\begin{equation}
\frac{\diff\sigma}{\diff m_{\pi\pi}} =\int_{-1}^1
\frac{|\bar{M}(E^2,s,\cos\theta)|^2
|\vec{k_3^\ast}||\vec{k_5}|}{128\pi^3 |\vec{k_1}|E^2}\diff
\cos\theta\,,\label{eq.pipimassdistribution}
\end{equation}
where $\vec{k_1}$ and $\vec{k_5}$ represent the 3-momenta of $e^\pm$ and $\phi$ in the
center-of-mass frame, respectively, and $\vec{k_3^\ast}$ denotes the 3-momenta of $\pi^\pm$ in the rest frame of the
$\pi\pi$ system. They are given as
\begin{equation}
|\vec{k_1}|=\frac{E}{2}\,, \quad
|\vec{k_3^\ast}|=\frac{1}{2}\sqrt{s-4m_\pi^2}\,, \quad
|\vec{k_5}|=\frac{1}{2E} \lambda^{1/2}\big(E^2,s,M_\phi^2\big) \,.
\end{equation}

The $\phi\pi$ invariant mass distribution of $e^+e^- \to \phi \pi^+\pi^-$ reads
\begin{equation}
\frac{\diff\sigma}{\diff m_{\phi\pi}} =\int_{s_-}^{s_+}
\frac{|\bar{M}(E^2,s,\cos\theta)|^2
|\vec{k_3^\ast}||\vec{k_5}|m_{\phi\pi}}{64\pi^3 |\vec{k_1}|E^2\kappa_{\pi}(s)\sqrt{s}}\diff
s\,,\label{eq.Jpsipimassdistribution}
\end{equation}
where
\begin{equation}
s_\pm=\frac{1}{4m_{\phi\pi}^2}\Big\{\big(E^2-M_\phi^2\big)^2 - \big[\lambda^{\frac{1}{2}}(E^2,m_{\phi\pi}^2,m_\pi^2)
\mp \lambda^{\frac{1}{2}}(m_{\phi\pi}^2,m_\pi^2,M_\phi^2)\big]^2\Big\}\,.
\end{equation}

\section{Phenomenological discussion}\label{pheno}

\subsection{Fitting to the BESIII data}

In this work we perform fits simultaneously by taking into account the experimental data sets of the $\pi\pi$ and $\phi\pi$
invariant mass distributions of $e^+e^- \to \phi
\pi\pi$ measured at a center-of-mass energy
$E=2.125\GeV$ by the BESIII
Collaboration~\cite{BESIII:2018rdg}.
The charged and neutral-pion final states data will be taken account of simultaneously.
\begin{figure}
\centering
\includegraphics[width=\linewidth]{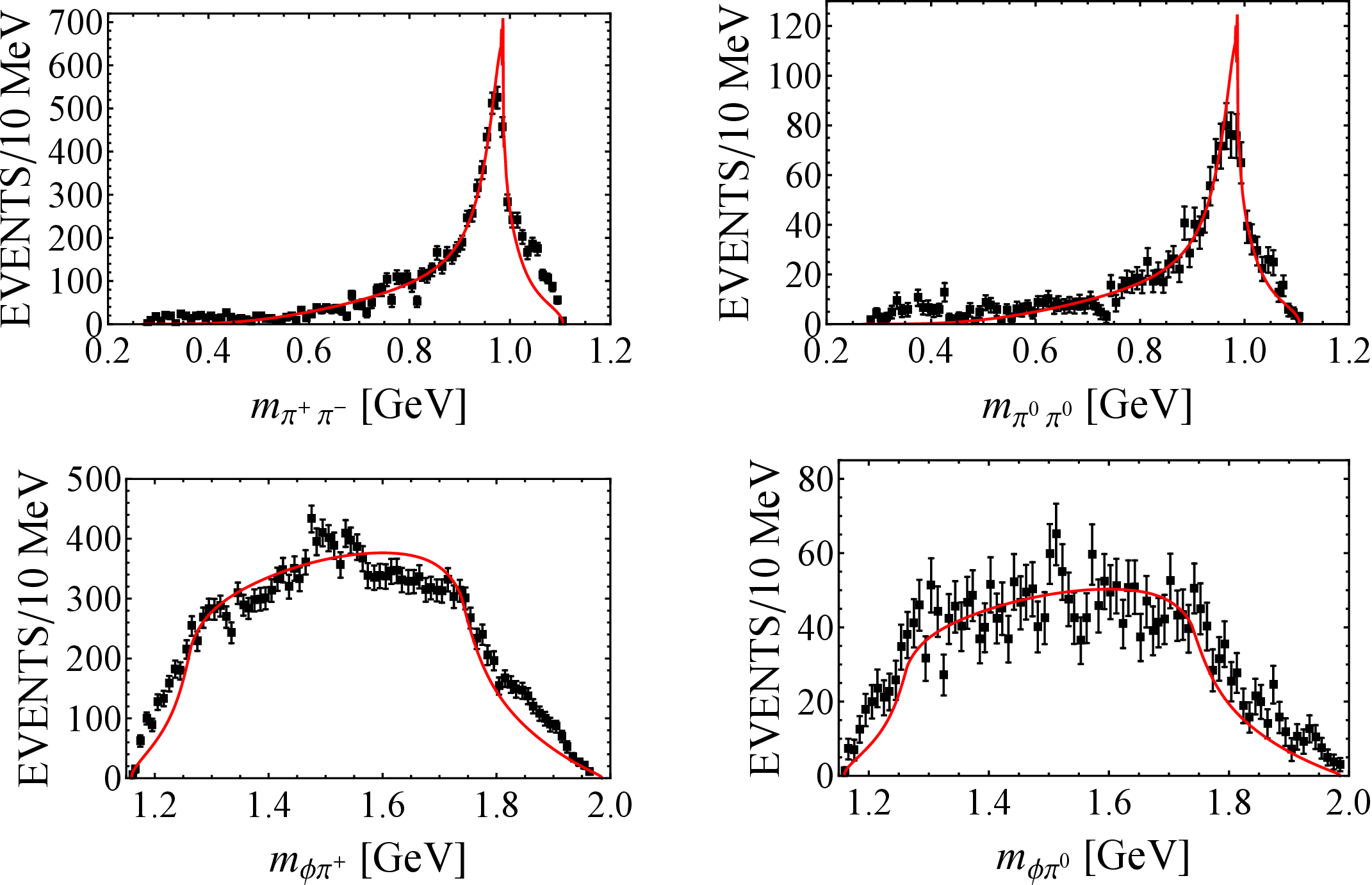}
\caption{Fit results of the $\pi\pi$ (top) and $\phi\pi$ (bottom) invariant mass spectra in
$e^+e^- \to \phi \pi^+\pi^-$ process. The charged- (left) and neutral-pion final states (right) data are taken into account Simultaneously.   The experimental data are taken from Ref.~\cite{BESIII:2018rdg}.
}\label{fig.MpipiMphipi}
\end{figure}

There are four free parameters in our fits: the combinations of the coupling constants in Eq.~\eqref{eq.ContactPi+KRaw} $(g_1-\sqrt{2}g_8)$ and $(h_1-\sqrt{2}h_8)$, and the two normalization factors $N_1$ and $N_2$ for the charged and neutral final states, respectively.
By performing the $\chi^2$ fit, we can determine the unknown combinations of
the resonance couplings:
\begin{eqnarray} \label{fit-result}
&&(g_1-\sqrt{2}g_8) =(-0.385\pm 0.005) \,,\qquad (h_1-\sqrt{2}h_8) =(3.120\pm 0.029)\,,
\end{eqnarray}
with $\chi^2/{\rm d.o.f }= 1583.9/(332-4)=4.83$.

In Fig.~\ref{fig.MpipiMphipi}, the fit results of the $\pi\pi$ and $\phi\pi$ mass spectra in $e^+e^- \to \phi \pi\pi$
are shown.
For the $\pi\pi$ mass spectra, one can see that the peak around 1 GeV due to the presence of the $f_0(980)$ is described well.
A small hump below 0.5 GeV cannot be reproduced in our scheme. Also note that there are differences between the shapes of the data for the modes with charged and neutral pions especially in the region close to the lower kinematical boundary and in the region close to
the upper kinematical boundary. These discrepancies contribute sizeably to the value of $\chi^2$.
For the $\phi\pi$ mass spectra, no $Z_s$ signal is observed. The data points below 1.25 GeV or above 1.8 GeV carrying small error bars contribute largely to the value of $\chi^2$. Note that in present scheme we only consider the leading chiral effective Lagrangian for the $\phi(2170)\phi PP$ contact couplings, and taking account of higher order chiral coupling terms may help to reduce the value of $\chi^2$. Nevertheless, since our theoretical predictions roughly agree with the experimental data, we discuss the fit results in more details.

Using the fitted coupling constants in Eq.~\eqref{fit-result}, we can calculate the cross sections ratio ${\sigma(e^+e^- \to \phi(2170)\to \phi K^+ K^-)}/{\sigma(e^+e^- \to \phi(2170)\to \phi\pi^+\pi^-)}$. At $\sqrt{s}=2.125$ GeV, our theoretical prediction is ${\sigma(e^+e^- \to \phi(2170)\to \phi K^+ K^-)}/{\sigma(e^+e^- \to \phi(2170)\to \phi\pi^+\pi^-)}=0.12 \pm 0.01$.
Note that the $\phi(2170)$ is very close to the $\phi KK$ threshold, and therefore the phase space of $e^+e^- \to \phi(2170)\to \phi K^+ K^-$ is much smaller than that of $e^+e^- \to \phi(2170)\to \phi\pi^+\pi^-$.
Using the experimental cross sections measured at the same energy point ${\sigma(e^+e^- \to \phi K^+ K^-)}=(70.6 \pm 7.2\pm 4.9)$ pb~\cite{BESIII:2019ebn} and ${\sigma(e^+e^- \to \phi\pi^+\pi^-)}=(436.2 \pm 6.4 \pm 30.1)$ pb~\cite{BESIII:2018rdg}, one obtains ${\sigma(e^+e^- \to \phi K^+ K^-)}/{\sigma(e^+e^- \to \phi\pi^+\pi^-)}=0.16 \pm 0.02$.
Note that due to the constructive or destructive interferences between/among different resonances and background, the multi-solution problem exists in using coherent contributions to fit the data, as have been pointed out in Refs.~\cite{0710.5627,0707.2541,0707.3699,Yuan:2009gd,1108.2760,1505.01509,1901.01394,Shen:2009mr}. As shown in Ref.~\cite{Shen:2009mr}, two solutions are found in the fit to the data of $e^+e^- \to \phi\pi^+\pi^-$ with two coherent Breit-Wigner functions. The products of the branching fraction of $\phi(2170)$ to $\phi\pi^+\pi^-$ and the $e^+e^-$ partial width in these two solutions are $68.9\pm7.0\pm3.4 $ eV/$c^2$ and $6.2\pm1.1\pm0.3$ eV/$c^2$, respectively, which differ with each other by one order. And thus it is questionable to attribute the experimental cross section of $e^+e^- \to \phi\pi^+\pi^-$ at 2.125 GeV totally to the $\phi(2170)$ intermediate state. Note that the experimental paper Ref.~\cite{BESIII:2019ebn} does not perform fit to the $e^+e^- \to \phi K^+ K^-$ cross section distribution to extract the parameters of $\phi(2170)$ and other resonances. When this kind of fits are available in the future, our results could be useful for selecting the physical solution.
On the other hand, if we assume that our estimation of the cross sections ratio ${\sigma(e^+e^- \to \phi(2170)\to \phi K^+ K^-)}/{\sigma(e^+e^- \to \phi(2170)\to \phi\pi^+\pi^-)}$ can be approximately extended to other energy points in the region around $\sqrt{s}=2.125$ GeV, we may infer that the peak in the $e^+e^- \to \phi K^+ K^-$ cross section distribution must also reflect in the $e^+e^- \to \phi \pi^+ \pi^-$ cross section distribution. Observe that there are only two obvious peaks, $\phi(2170)$ and $X(2400)$, in the experimental $e^+e^- \to \phi \pi^+ \pi^-$ cross section distribution in the region of [2.0, 2.6] GeV, and the $X(2400)$ affects the resonance parameter of $\phi(2170)$ only moderately due to its width is only about 100 MeV~\cite{Shen:2009mr}. One may attribute the experimental cross section ${\sigma(e^+e^- \to \phi K^+ K^-)}$ at 2.125 GeV mainly to the $\phi(2170)$ intermediate state. Therefore using our theoretical prediction of cross sections ratio given above one can obtains that the lager solution of the product of the branching fraction of $\phi(2170)$ to $\phi\pi^+\pi^-$ and the $e^+e^-$ partial width $68.9\pm7.0\pm3.4 $ eV/$c^2$ is preferred, since in this solution the peak due to the $\phi(2170)$ intermediate state is higher than the cross section data at $\sqrt{s}=2.125$ GeV, as shown in Fig.~3 of Ref.~\cite{Shen:2009mr}. Also one may conclude that above the kinematical threshold of $\phi K\bar{K}$ the mechanism $e^+e^- \to ... \to \phi K^+ K^-$ with the kaons rescattering to a pion pair may be an important contribution to $e^+e^- \to \phi\pi^+\pi^-$.
In addition, using our fitted results we can calculate the ratio of the cross sections $\sigma (e^+e^- \to \phi(2170)
\to \phi \pi^0\pi^0)/\sigma(e^+e^- \to \phi(2170)
\to \phi \pi^+\pi^-)=0.51\pm 0.02$, which agree with the experimental ratio given in Ref.~\cite{BESIII:2018rdg}, $0.54\pm 0.6$, within error bar.

\section{Conclusions}
\label{conclu}

We have used dispersion theory to study the processes
$e^+e^-\to \phi(2170) \to \phi \pi\pi(K\bar{K})$. The strong FSI between
two pseudoscalar mesons has been considered in a model-independent way, and
the leading chiral amplitude acts as the subtraction function in the
Omn\`es solution.
Through fitting to the data of the $\pi\pi$ and $\phi\pi$
invariant mass spectra of $e^+e^-\to \phi(2170) \to \phi \pi\pi$,
the couplings of the $\phi(2170) \phi PP$ vertex are determined.
We give the prediction of the cross sections ratio ${\sigma(e^+e^- \to \phi(2170)\to \phi K^+ K^-)}/{\sigma(e^+e^- \to \phi(2170)\to \phi\pi^+\pi^-)}$, and the result could be useful for
selecting the physical solution when the fit to the $e^+e^- \to \phi K^+ K^-$ cross section distribution is available in the future.
Our findings suggest that above the kinematical threshold of $\phi K\bar{K}$, the mechanism $e^+e^- \to \phi K^+ K^-$ with the kaons rescattering to a pion pair plays an important role in the $e^+e^- \to \phi\pi^+\pi^-$ transition.

\section*{Acknowledgments}

This work is supported in part by the Fundamental Research Funds
for the Central Universities under Grants No.~FRF-BR-19-001A, by the National Natural Science Foundation of China (NSFC) under Grants No.~11975028, No.~11974043.

\end{document}